\documentclass[lettersize,journal]{IEEEtran}
\usepackage{amsmath,amsfonts}
\usepackage{amssymb}
\usepackage{algorithm}
\usepackage{array}
\usepackage[caption=false,font=normalsize,labelfont=sf,textfont=sf]{subfig}
\usepackage{textcomp}
\usepackage{stfloats}
\usepackage{url}
\usepackage{verbatim}
\usepackage{graphicx}
\usepackage{algpseudocode}
\usepackage{hyperref}
\hypersetup{
    colorlinks=true,
    linkcolor=blue,
    citecolor=blue,
    urlcolor=blue}
\begin{document}

\title{Innovation-Domain Decision-Directed Phase Tracking \\ for Wiener Phase Noise in Fast Rayleigh Fading}

\author{Ura Klongklaew, Nithiroth Pornsuwancharoen, and Phichai Youplao~
\thanks{This is the accepted version of the article published in IEEE Wireless Communications Letters. The final version is available at https://doi.org/10.1109/LWC.2026.3696948. © 2026 IEEE.}
}



\maketitle

\begin{abstract}
This letter proposes an innovation-domain decision-directed phase tracking (ID-DDPT) architecture for coherent detection over Rayleigh fading channels with temporally correlated phase evolution and Wiener phase noise. By reformulating phase tracking into the innovation domain, replacing the unbounded cumulative phase with its stationary increments, the proposed method converts a non-stationary estimation problem into a stable low-complexity filtering problem. A closed-form expression for the steady-state residual phase error variance is derived under the locked-regime assumption, along with an analytical optimal smoothing factor. Modeling the residual phase distortion as an effective signal-to-noise ratio (SNR) attenuation yields a tractable bit error rate (BER) approximation for BPSK over Rayleigh fading. A first-order error-propagation analysis further characterizes the impact of decision errors and provides insight into the onset of cycle slips. Simulation results demonstrate that ID-DDPT outperforms DBPSK and a complexity-equivalent scalar Kalman tracker (SKT), achieving near-coherent performance with $\mathcal{O}(1)$ per-symbol complexity and minimal pilot overhead.
\end{abstract}


\section{Introduction}

\IEEEPARstart{O}{scillator} phase noise is a fundamental impairment in coherent wireless receivers, particularly in low-cost free-running oscillator designs and high-mobility scenarios. Random phase fluctuations corrupt the carrier phase reference and degrade coherent detection performance if left uncompensated; their impact has accordingly attracted extensive study in both sub-6~GHz and millimeter-wave systems \cite{Chang2025, Demir2000}.

The free-running oscillator phase is canonically modeled as a discrete-time Wiener process with Gaussian innovations \cite{Masoumi2026}. This formulation captures the unbounded variance growth of the cumulative phase $\theta_k$, which poses a fundamental challenge for coherent detection, particularly under the joint impairments of Rayleigh fading amplitude and phase fluctuations \cite{Fodor2022, Li2023}. Classical phase-locked loop (PLL) techniques estimate $\theta_k$ directly but can suffer from instability or slow convergence when phase innovations are large \cite{Valls2017, Ripani2024}. Noncoherent or differential detection sidesteps phase estimation entirely yet incurs an inherent SNR penalty in fading channels \cite{Simon2002, Gunasekara2025}. 

More sophisticated Bayesian or iterative state-space approaches improve phase estimation accuracy at the cost of increased computational complexity through Kalman-based recursions or iterative message passing \cite{Colavolpe2005, Mohammadian2021, Yu2026}. Recent coding and modulation advances, such as protograph LDPC and shaped index modulation for OAM systems~\cite{Yang2024}, improve reliability under ideal synchronization but remain complementary to phase tracking. Generalized fading models, including the $\lambda$-$\kappa$-$\mu$, double $\lambda$-$\kappa$-$\mu$, $\lambda$-$\kappa$-$\mu$ shadowed, and $\alpha$-$\eta$-$\mu$/IGA distributions, have also been investigated for diverse propagation environments \cite{Wang2024, Liu2025, Xu2026}. While these models offer superior fidelity, Rayleigh fading remains a standard baseline for tractable analytical characterization. Nevertheless, robust, low-complexity tracking under joint phase noise and fading remains a persistent challenge.

A key observation motivates this reformulation: in contrast to conventional phase tracking methods (e.g., PLL and Kalman filtering) that operate on the non-stationary cumulative phase $\theta_k$, the phase increment $\Delta_k = \theta_k - \theta_{k-1}$ is stationary with bounded variance. This suggests reformulating phase tracking in the innovation domain, transforming an unbounded random-walk estimation problem into a stationary filtering problem, thereby avoiding direct filtering of the divergent cumulative phase trajectory and enabling closed-form steady-state analysis with $\mathcal{O}(1)$ complexity.

Based on this insight, this letter proposes a low-complexity innovation-domain decision-directed phase tracking (ID-DDPT) architecture. The principal contributions are fourfold: (i) a clarified channel-phase model that separates the Rayleigh fading phase from the Wiener oscillator dynamics and incorporates it as an effective perturbation; (ii) closed-form expressions for the steady-state residual phase variance and the optimal exponential moving average (EMA) smoothing factor; (iii) a first-order error-propagation analysis that quantifies the gap between the ideal lower bound and the practical decision-directed loop; and (iv) simulation comparisons against DBPSK and a scalar Kalman tracker (SKT) of identical $\mathcal{O}(1)$ complexity, demonstrating near-coherent bit error rate (BER) performance under moderate phase noise.


\section{System Model}
\label{sec:model}

Consider a discrete-time single-input single-output BPSK link over a frequency-flat Rayleigh fading channel impaired by oscillator phase noise. The ID-DDPT receiver requires only a single initial pilot to resolve the absolute phase ambiguity.

At symbol interval $k \in \mathbb{Z}$, a bit $b_k \in \{0,1\}$ is mapped to 
\begin{equation}
x_k = (2b_k - 1)\sqrt{E_s} \in \{\pm \sqrt{E_s}\}.
\end{equation}
The received complex baseband signal is given by
\begin{equation}
r_k = a_k x_k e^{j\phi_k} + n_k,
\label{eq:rk_model}
\end{equation}
where $a_k \ge 0$ denotes the Rayleigh fading amplitude, $\phi_k$ is the composite carrier phase, and $n_k \sim \mathcal{CN}(0, N_0)$ is additive white Gaussian noise (AWGN). To account for the combined effects of oscillator impairments and channel-induced phase variations, the composite phase $\phi_k$ is further decomposed as
\begin{equation}
\phi_k = \theta_k + \psi_k,
\end{equation}
where $\theta_k$ represents oscillator phase noise and $\psi_k$ denotes the fading-induced phase.

The oscillator phase $\theta_k$ follows a discrete-time Wiener process
\begin{equation}
\theta_k = \theta_{k-1} + \Delta_k,\quad \Delta_k \sim \mathcal{N}(0,\sigma_\Delta^2),
\end{equation}
which models accumulated phase drift with independent Gaussian innovations.

The fading-induced phase $\psi_k$ is modeled as a temporally correlated stochastic process consistent with Clarke/Jakes-type fading, characterized by Doppler spread $f_D$ and coherence time $T_c \approx 1/f_D$. In this work, \textit{fast fading} refers to scenarios with non-negligible Doppler spread but satisfying the finite-coherence condition $T_s \ll T_c$, where $T_s$ is the symbol duration. Under this condition, adjacent channel realizations remain statistically correlated, and the fading-induced increment $\psi_k - \psi_{k-1}$ exhibits bounded variance governed by the Doppler dynamics.

In contrast, when $T_s \gtrsim T_c$, the channel becomes effectively uncorrelated across symbols, and incremental phase tracking is no longer reliable due to the absence of exploitable temporal structure in successive observations.

Under the finite-coherence regime $T_s \ll T_c$, the composite phase increment can be expressed as
\begin{equation}
\phi_k - \phi_{k-1} = \Delta_k + (\psi_k - \psi_{k-1}),
\end{equation}
where $\Delta_k$ captures oscillator phase innovations and $(\psi_k - \psi_{k-1})$ represents a Doppler-induced perturbation.

It is important to emphasize that the composite phase does not strictly follow a Wiener recursion due to the presence of the fading component. However, the proposed receiver operates on phase increments, which remain well-defined under finite-coherence channels and preserve the temporal structure required for reliable tracking.

This formulation naturally leads to two operating regimes: a locked regime ($T_s \ll T_c$), where stable tracking is achievable, and a slip-dominated regime ($T_s \gtrsim T_c$), where phase continuity cannot be maintained. This regime-based interpretation is consistent with the performance behavior observed in Section~\ref{results_discussion}. For notational clarity, the average channel power is normalized as $\mathbb{E}[a_k^2] = 1$, yielding an average signal-to-noise ratio $\bar{\gamma} = E_s / N_0$.

\section{Proposed Innovation-Domain Phase Tracking}
\label{sec:detector}

\begin{figure}[!t]
\centering
\includegraphics[width=3.5in]{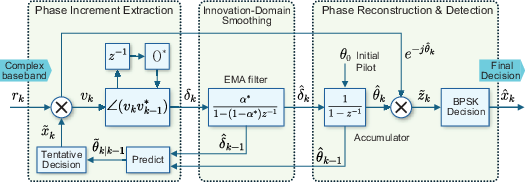}
\caption{System model and the proposed ID-DDPT receiver architecture.}
\label{fig:sys_model}
\end{figure}

As depicted in Fig.~\ref{fig:sys_model}, the receiver operates in three stages summarized in Algorithm~\ref{alg:id_ddpt}: (i) phase-increment extraction via decision-directed data wipe-off; (ii) innovation-domain EMA smoothing; and (iii) phase reconstruction and coherent detection.

\begin{algorithm}[!t]
\caption{ID-DDPT}
\label{alg:id_ddpt}
\begin{algorithmic}[1]
\renewcommand{\algorithmicrequire}{\textbf{Input:}}
\renewcommand{\algorithmicensure}{\textbf{Output:}}

\State \textbf{Initialization:} 
\State \quad Acquire $\hat{\theta}_{0}$ from an initial pilot symbol
\State \quad Compute innovation-to-noise ratio: $K \leftarrow 2\bar{\gamma}\sigma_{\Delta}^{2}$$^\dagger$
\State \quad Optimal smoothing factor: $\alpha^{\star} \leftarrow \frac{\sqrt{K^{2} + 4K} - K}{2}$
\State \quad Set $\hat{\delta}_{0} \leftarrow 0$ and $v_{0} \leftarrow \sqrt{E_{s}}e^{j\hat{\theta}_{0}}$

\For{$k = 1, 2, \dots$}
    \State Predict phase: $\tilde{\theta}_{k|k-1} \leftarrow \hat{\theta}_{k-1} + \hat{\delta}_{k-1}$
    \State Tentative decision: $\tilde{x}_{k} \leftarrow \text{sign}(\mathfrak{R}\{r_{k}e^{-j\tilde{\theta}_{k|k-1}}\})\sqrt{E_{s}}$
    \State Data wipe-off: $v_{k} \leftarrow r_{k}\tilde{x}_{k}$
    \State Extract increment: $\delta_{k} \leftarrow \angle(v_{k}v_{k-1}^{*})$
    \State $\hat{\delta}_{k} \leftarrow \alpha^{\star}\delta_{k} + (1-\alpha^{\star})\hat{\delta}_{k-1}$
    \State Update phase: $\hat{\theta}_{k} \leftarrow \hat{\theta}_{k-1} + \hat{\delta}_{k}$
    \State Final decision: $\hat{x}_{k} \leftarrow \text{sign}(\mathfrak{R}\{r_{k}e^{-j\hat{\theta}_{k}}\})\sqrt{E_{s}}$
\EndFor
\end{algorithmic}
{\footnotesize $^\dagger$Approximates $\sigma^2_w \approx 1/(2\bar{\gamma})$, neglecting $\sigma^2_\psi$.}
\end{algorithm}

\subsection{Stage 1: Data Wipe-off and Phase-Increment Extraction}

Given the predicted phase $\tilde{\theta}_{k|k-1} = \hat{\theta}_{k-1} + \hat{\delta}_{k-1}$, a tentative decision is formed as $\tilde{x}_{k} = \text{sign}(\Re\{r_{k} e^{-j\tilde{\theta}_{k|k-1}}\})\sqrt{E_{s}}$. Data modulation is wiped off to obtain $v_k = r_k \tilde{X}_k$, from which the instantaneous phase increment is extracted via the conjugate product
\begin{equation}
  \delta_k \approx \Delta_k + \varepsilon_k,
  \label{eq:delta_k}
\end{equation}
where $\Delta_k$ is the true Wiener innovation and $\varepsilon_k$ is an additive perturbation. This approximation assumes correct tentative symbol decisions (i.e., perfect data wipe-off), and therefore characterizes the best-case steady-state performance of the decision-directed loop under the condition of correct decisions. Although the Rayleigh fading amplitude $a_k$ is modeled as i.i.d. across symbols, the fading-induced phase $\psi_k$ remains temporally correlated under the finite-coherence condition ($T_s \ll T_c$). The conjugate-product operation therefore exploits this residual phase continuity, while the Doppler-induced increment ($\psi_k - \psi_{k-1}$) and additive noise contribute as a superimposed perturbation $\varepsilon_k$ mitigated by the subsequent EMA smoothing stage. At moderate-to-high SNR with correct tentative decisions, $\varepsilon_k$ receives three statistically independent contributions: (i) additive Gaussian noise with variance $\sigma_n^2 \approx 1 / (2 \bar\gamma)$, obtained by second-order Taylor expansion \cite{Proakis2008}; (ii) the fading-phase difference ($\psi_k - \psi_{k-1}$), which for i.i.d. $\psi_k$ uniform on $[0,2\pi)$ is a zero-mean random variable with variance $\sigma_\psi^2 \le 2 \text{ rad}^2$; and (iii) an amplitude-ratio perturbation from consecutive Rayleigh envelopes, absorbed into (ii) for tractability.

For analytical tractability, $\varepsilon_k$ is modeled as a zero-mean random variable with total variance $\sigma_w^2 = \sigma_n^2 + \sigma_\psi^2$. Although the fading-phase difference ($\psi_k - \psi_{k-1}$) is not Gaussian in general, it is approximated via moment matching for variance-based BER analysis, and this approximation yields accurate performance predictions as validated by simulation. The bound $\sigma_\psi^2 \le 2$ corresponds to the worst-case i.i.d. uncorrelated fading scenario; in practice, non-zero temporal correlation (moderate Doppler) reduces $\sigma_\psi^2$ substantially. The total effective measurement noise variance is thus
\begin{equation}
  \sigma_w^2 = \sigma_n^2 + \sigma_\psi^2 = \frac{1}{2 \bar\gamma} + \sigma_\psi^2.
  \label{eq:sigma_w}
\end{equation}
This treatment represents a local approximation enabling closed-form steady-state analysis. Although the Doppler-induced increment process $(\psi_k-\psi_{k-1})$ remains temporally correlated under finite-coherence fading, its correlation is sufficiently weak in the locked regime ($T_s \ll T_c$) that its dominant effect can be captured through the effective innovation variance $\sigma_w^2$. Residual temporal correlation may slightly shift the optimal smoothing factor and affect transient tracking behavior in practical correlated fading scenarios. Unlike the cumulative phase $\theta_k$, the increment variance $\sigma_\delta^2 = \sigma_\Delta^2 + \sigma_w^2$ remains bounded and time-invariant, enabling stable time-invariant filtering under finite-coherence fading conditions.

\subsection{Stage 2: Innovation-Domain EMA Smoothing}

A first-order exponential moving average filter is applied to the extracted increments:
\begin{equation}
  \hat{\delta}_k = \alpha \delta_k + (1 - \alpha)\hat{\delta}_{k-1}, \quad 0 < \alpha \le 1.
  \label{eq:sigma_k}
\end{equation}

The transfer function $H(z) = \alpha /\left[ 1-(1-\alpha) z^{-1} \right]$ is unconditionally stable for all $\alpha \in (0, 1]$. The extracted increment sequence $\delta_k$ is modeled as locally white in the locked regime, where decision errors are sparse and temporal correlation remains weak. Under this approximation, the output variance of the EMA filter is
\begin{equation}
  \text{Var}(\hat{\delta}_k) = \frac{\alpha}{2 - \alpha} \sigma^2_\delta.
  \label{eq:var_delta_k}
\end{equation}

\subsection{Stage 3: Phase Reconstruction and Detection}

Smoothed increments are accumulated as $\hat{\theta}_k = \hat{\theta}_{k-1} + \hat{\delta}_k$. The residual phase error $\varphi_k = \theta_k - \hat{\theta}_k$ governs system performance and is analyzed in Section~\ref{performance}. The final compensated signal $\tilde{z}_k = r_k e^{-j\hat{\theta}_k} \approx a_k x_k e^{j\varphi_k} + n_k e^{-j\hat{\theta}_k}$ is passed through a standard BPSK decision. The complete per-symbol procedure ($\mathcal{O}(1)$ complexity) is given in Algorithm~\ref{alg:id_ddpt}.

\section{Performance Analysis}
\label{performance}

\subsection{Residual Phase Variance}

The ID-DDPT estimator applies EMA smoothing $H(z)$ to the noisy increment $\delta_k = \Delta_k + \varepsilon_k$, then accumulates the smoothed output to form $\hat{\theta}_k$. Because $\Delta_k$ and $\varepsilon_k$ are independent, the residual phase error $\varphi_k = \theta_k - \hat{\theta}_k$ decomposes into two independent contributions and its variance is the sum of their individual variances. The first contribution arises from the untracked portion of the Wiener innovation. In the $z$-domain, the true phase is $\Theta(z) = \Delta(z)/(1-z^{-1})$ and the $\Delta$-driven part of the estimate is $H(z)\Delta(z)/(1-z^{-1})$, so the error transfer function from $\Delta_k$ to $\varphi_k$ is
\begin{equation}
E_{\Delta}(z) = \frac{1-H(z)}{1-z^{-1}} = \frac{1-\alpha}{1-(1-\alpha)z^{-1}},
\label{eq:e_Delta}
\end{equation}
which is stable for all $\alpha \in (0, 1]$, since the pole of $E_{\Delta}(z)$ lies at $z = (1-\alpha)$ and satisfies $|1-\alpha| < 1$. Applying Parseval's theorem to $|E_{\Delta}(e^{j\omega})|^2$ and evaluating the resulting contour integral gives the $\Delta$-contribution to the residual variance:
\begin{equation}
\text{Var}_{\Delta}(\varphi_k) = \frac{(1-\alpha)^{2}}{\alpha(2-\alpha)} \sigma_{\Delta}^{2}.
\end{equation}
The second contribution arises from the measurement noise $\varepsilon_k$. The EMA filter smooths $\varepsilon_k$ before phase reconstruction, yielding the variance contribution
\begin{equation}
\text{Var}_{\varepsilon}(\varphi_k)
=
\frac{\alpha}{2-\alpha}\sigma_{w}^{2}.
\end{equation}
Summing the two independent contributions gives the closed-form steady-state residual phase variance:
\begin{equation}
\sigma_{\varphi}^{2}(\alpha)
=
\frac{(1-\alpha)^{2}}{\alpha(2-\alpha)} \sigma_{\Delta}^{2}
+
\frac{\alpha}{2-\alpha} \sigma_{w}^{2}.
\label{eq:sigma_varphi}
\end{equation}
This expression remains bounded for all $\alpha \in (0,1]$ despite the unbounded variance growth of $\theta_k$, since the innovation-domain formulation filters phase increments rather than the cumulative phase trajectory directly.

In practical correlated fading channels, residual temporal correlation in $(\psi_k-\psi_{k-1})$ may slightly shift the optimal $\alpha$ from the closed-form solution, as validated by the close agreement between analysis and simulation in Section~\ref{results_discussion}.

\subsection{Optimal Smoothing Factor}

Minimizing~\eqref{eq:sigma_varphi} over $\alpha$ yields the quadratic $\sigma_{w}^{2} \alpha^{2} + \sigma_{\Delta}^{2} \alpha - \sigma_{\Delta}^{2} = 0$. Defining the innovation-to-noise ratio $K = \sigma_{\Delta}^{2}/\sigma_{w}^{2} \approx 2\bar{\gamma}\sigma_{\Delta}^{2}$ and solving for the positive root gives
\begin{equation}
\alpha^{\star} = \frac{\sqrt{K^{2} + 4K} - K}{2}.
\label{eq:alpha_star}
\end{equation}
As $K \rightarrow 0$ (low SNR or stable oscillator), $\alpha^{\star} \rightarrow 0$, emphasizing historical averaging to suppress measurement noise. As $K \rightarrow \infty$ (high SNR or severe phase noise), $\alpha^{\star} \rightarrow 1$, enabling tight tracking of rapid phase increments. This closed-form result allows the receiver to self-configure based solely on the operating SNR and oscillator specifications without empirical tuning.

Residual temporal correlation in the Doppler-induced increment process may slightly shift the optimal smoothing factor in practical correlated fading channels. The derived $\alpha^\star$ therefore represents a tractable analytical approximation under the locally white innovation model.

\begin{figure*}[!t]
\centering
\includegraphics[width=7.0in]{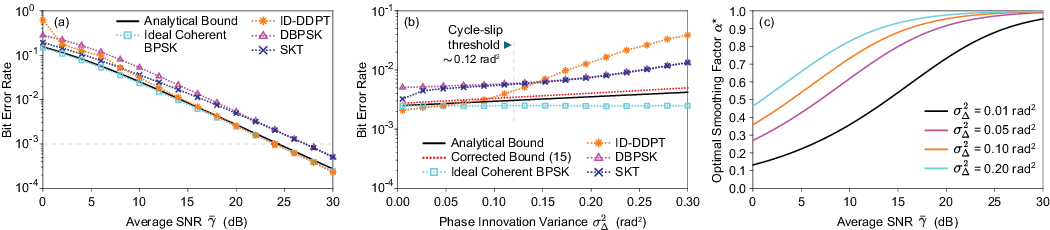}
\caption{Performance of the proposed ID-DDPT detector over
correlated Rayleigh flat-fading channels with Wiener phase noise.
(a) BER versus average SNR ($\sigma_\Delta^2 = 0.05$\,rad$^2$,
$\alpha = 0.3$), where ID-DDPT achieves a $2.5$--$3.0$\,dB gain
over DBPSK and Scalar Kalman tracker (SKT) at BER $10^{-3}$ and approaches ideal coherent BPSK.
(b) BER versus phase innovation variance
($\overline{\gamma}=20$\,dB, $\alpha=0.3$), showing accurate
tracking for $\sigma_\Delta^2 \le 0.12$\,rad$^2$. Solid lines
denote analytical bounds, and markers indicate Monte Carlo
results ($10^7$ symbols). (c) Optimal EMA smoothing factor
$\alpha^\star$ versus average SNR.}
\label{fig:results}
\end{figure*}

\subsection{Error Propagation and Cycle-Slip Threshold}
\label{sec:error_prop}

The analytical BER in Section~\ref{ber_approx} assumes perfect data wipe-off ($\tilde{x}_k = x_k$). In the practical decision-directed (DD) loop, tentative decision errors introduce spurious $\pm\pi$ phase jumps with probability $P_e^{(t)}$. These errors inflate the effective innovation variance. A first-order phenomenological model captures the dominant impact of decision errors on the effective innovation variance as
\begin{equation}
\sigma_{\Delta,\text{eff}}^{2} = \sigma_{\Delta}^{2} + \pi^{2} \times 2P_{e}^{(t)}(1 - P_{e}^{(t)}),
\label{eq:sigma_Delta}
\end{equation}
where $P_e^{(t)}$ is computed from~\eqref{eq:ber_for_bpsk} using the current $\sigma_{\varphi}^{2}$. Substituting~\eqref{eq:sigma_Delta} into~\eqref{eq:sigma_varphi} and iterating to a fixed point yields a corrected variance $\sigma_{\varphi,\text{corr}}^{2}$ that tracks simulated performance up to the cycle-slip threshold; beyond it, the iteration diverges, indicating error propagation. Using $\sigma_{\varphi,\text{eff}}^{2} \le 0.3\,\text{rad}^2$ as the limit of the Gaussian approximation (Section~\ref{ber_approx}), the threshold is $\sigma_{\Delta}^{2} \approx 0.12\,\text{rad}^2$ at $\bar{\gamma}=20$ dB, consistent with simulations.

\subsection{Closed-Form BER Approximation}
\label{ber_approx}

Conditioned on fading SNR $\gamma$, the residual phase $\varphi_k \sim \mathcal{N}(0, \sigma_\varphi^2)$. Using the moment-matched Gaussian approximation, $\mathbb{E}[\cos \varphi_k] = e^{-\sigma_\varphi^2/2}$ \cite{Proakis2008}. The conditional BER for coherent BPSK is
\begin{equation}
P_e(\gamma) \approx Q\left(\sqrt{2\gamma} e^{-\sigma_\varphi^2/2}\right).
\label{eq:ber_for_bpsk}
\end{equation}
Averaging over the Rayleigh fading distribution $f_\gamma(\gamma) = (1/\bar{\gamma}) e^{-\gamma/\bar{\gamma}}$ and applying the standard closed-form result for coherent BPSK over Rayleigh fading \cite{Simon2002} gives
\begin{equation}
P_e \approx \frac{1}{2} \left(1 - \sqrt{\frac{\bar{\gamma} e^{-\sigma_\varphi^2}}{1 + \bar{\gamma} e^{-\sigma_\varphi^2}}} \right).
\label{eq:ber_approx}
\end{equation}
As $\bar{\gamma} \rightarrow \infty$, a Taylor expansion of~\eqref{eq:ber_approx} yields $P_e \sim e^{\sigma_\varphi^2}/(4\bar{\gamma})$, preserving first-order diversity for all finite $\sigma_\varphi^2$. Innovation-domain smoothing thus preserves the diversity order of coherent BPSK over Rayleigh fading while introducing only a coding-gain penalty $e^{-\sigma_\varphi^2}$.

\section{Numerical Results and Discussion}
\label{results_discussion}

Monte Carlo simulations use $10^7$ symbols per SNR point. ID-DDPT is evaluated against: (i) ideal coherent BPSK; (ii) conventional DBPSK; and (iii) a scalar Kalman tracker operating on the cumulative phase $\theta_k$ with a 1-D state-space model (process noise $\sigma_\Delta^2$, observation noise $1/(2\bar{\gamma})$), implemented using a steady-state Kalman gain and therefore yielding the same $\mathcal{O}(1)$ per-symbol complexity as ID-DDPT; classical PLL-based methods are omitted due to the difficulty of fair tuning under fast phase variations and fading conditions. The Kalman tracker is implemented in its scalar steady-state form without parameter adaptation, representing a baseline low-complexity state-space approach. The analytical curve~\eqref{eq:ber_approx} and the corrected bound from~\eqref{eq:sigma_Delta}--\eqref{eq:ber_approx} are overlaid on each figure.

Fig.~\ref{fig:results}(a) plots BER versus average SNR $\bar{\gamma} \in [0, 30]$~dB for $\sigma_{\Delta}^{2} = 0.05 \,\mathrm{rad}^2$. The proposed ID-DDPT consistently outperforms DBPSK across the entire SNR range, achieving an approximate gain of 2.5--3.0~dB at BER $= 10^{-3}$, and operates close to ideal coherent BPSK with a gap typically within approximately 1~dB. The SKT also improves upon DBPSK but exhibits a slightly larger performance gap relative to ID-DDPT, particularly in the low-to-medium SNR regime. This difference is primarily due to the underlying estimation domain: while SKT operates on the unbounded cumulative phase trajectory, the proposed ID-DDPT performs filtering on bounded and stationary phase increments in the innovation domain, thereby avoiding variance accumulation and improving robustness to phase uncertainty. All schemes exhibit the characteristic $1/\bar{\gamma}$ decay, demonstrating preservation of first-order diversity under Wiener phase noise.

Fig.~\ref{fig:results}(b) illustrates BER versus phase innovation variance $\sigma_{\Delta}^{2} \in [0.005, 0.30] \,\mathrm{rad}^2$ at $\bar{\gamma} = 20$~dB. The solid curve denotes the analytical bound~\eqref{eq:ber_approx}, while the red dotted curve represents the corrected bound incorporating error propagation via~\eqref{eq:sigma_Delta}. For small-to-moderate innovation variances ($\sigma_{\Delta}^{2} \lesssim 0.1$), the simulated ID-DDPT closely follows the corrected bound and outperforms both DBPSK and the SKT. As $\sigma_{\Delta}^{2}$ increases, performance degrades due to growing phase uncertainty. Beyond an empirical threshold of approximately $\sigma_{\Delta}^{2} \approx 0.12 \,\mathrm{rad}^2$, cycle slips become more frequent, leading to deviation from the analytical prediction and a rapid BER increase for ID-DDPT; this threshold is marked by a vertical dashed line in the figure, consistent with the cycle-slip onset predicted in Section~\ref{sec:error_prop}. In this regime, the performance gap with DBPSK diminishes, and DBPSK may become more robust due to its noncoherent nature. The SKT shows a similar degradation trend. These results illustrate the transition from a noise-limited regime, where the analytical model is accurate, to a slip-dominated regime governed by error propagation.

The deviation between analytical and simulation results can be explained within a unified regime-based framework. The analytical results in Section~\ref{sec:detector} assume perfect data wipe-off and ideal phase tracking, and thus represent a lower bound under stable operation. In the practical DD receiver, residual phase errors may propagate through feedback, and when the phase innovation variance increases or the SNR decreases, decision errors become more frequent, leading to cycle slips. The resulting performance is governed by the effective phase innovation variance $\sigma_{\phi}^2 = \sigma_\Delta^2 + \mathrm{Var}(\psi_k - \psi_{k-1})$, where $\sigma_\Delta^2$ denotes oscillator phase noise and $\mathrm{Var}(\psi_k - \psi_{k-1})$ captures Doppler-induced fading under coherence time $T_c \approx 1/f_D$.

Two operating regimes are observed. In the locked regime ($T_s \ll T_c$), temporal correlation enables reliable conjugate-product phase increment extraction, and the analytical bound closely matches simulation. When $T_s \gtrsim T_c$, progressive loss of temporal correlation increases decision errors and cycle slips, causing deviation from the steady-state analysis and transition to a slip-dominated regime. Thus, although the system targets fast Rayleigh fading, it operates under finite-coherence conditions rather than an i.i.d. fading assumption. The analytical expressions therefore represent lower bounds valid under stable tracking conditions.

Fig.~\ref{fig:results}(c) depicts the optimal smoothing factor $\alpha^{\star}$ from~\eqref{eq:alpha_star} as a function of $\bar{\gamma}$ for several values of $\sigma_{\Delta}^{2}$. The close agreement between analytical and simulation results in Fig.~\ref{fig:results}(a)--(b) supports the underlying assumptions, including the local whiteness approximation of the extracted increment sequence. This supports the use of the closed-form variance expression and the derived $\alpha^\star$ as accurate approximations in the locked regime despite residual Doppler-induced temporal correlation.

The proposed ID-DDPT framework focuses on BPSK to enable clear analytical characterization. The underlying innovation-domain structure is not restricted to BPSK. For QPSK, modulation can be removed via standard $M$-th power transformations or decision-directed wipe-off, after which the same increment-domain processing applies.

For higher-order QAM, extension requires incorporation of decision reliability to mitigate increased error propagation, e.g., via soft decisions or reliability-weighted updates, while the core innovation-domain filtering remains unchanged. A comprehensive treatment of higher-order modulation is beyond the scope of this letter and is left for future work, including the development of reliability-aware and soft-decision-based extensions of the innovation-domain tracking framework.

\section{Conclusion}

This letter proposed an innovation-domain decision-directed phase tracking architecture for coherent detection over Rayleigh fading channels impaired by Wiener phase noise. By operating on bounded phase increments rather than the cumulative phase trajectory, the proposed framework enables stable low-complexity tracking with an analytical characterization. Closed-form expressions for the residual phase variance, the optimal EMA smoothing factor, and the average BER were derived, together with a first-order error-propagation model capturing the onset of cycle slips. Simulation results showed that, for moderate innovation rates ($\sigma_{\Delta}^{2} \le 0.12 \,\text{rad}^2$), ID-DDPT provides superior BER performance over both DBPSK and a complexity-equivalent scalar Kalman tracker, approaching coherent detection with $\mathcal{O}(1)$ per-symbol complexity and a single initial pilot.

\end{document}